# Machine learning potential for predicting thermal conductivity of θ-phase and amorphous Tantalum Nitride

Zhicheng Zong[1,2], Yangjun Qin[1,2], Jiahong Zhan[1,2], Haisheng Fang[1], Nuo Yang[2*]

1. School of Energy and Power Engineering, Huazhong University of Science and Technology, Wuhan 430074, China.
2. School of Science, National University of Defense Technology, Changsha, 410073, China;

*Corresponding email: nuo@nudt.edu.cn (N. Yang)

**Abstract**

Tantalum nitride (TaN) has attracted considerable attention due to its unique electronic and thermal properties, high thermal conductivity, and applications in electronic components. However, for the θ-phase of TaN, significant discrepancies exist between previous experimental measurements and theoretical predictions. In this study, deep potential models for TaN in both the θ-phase and amorphous phase were developed and employed in molecular dynamics simulations to investigate the thermal conductivities of bulk and nanofilms. The simulation results were compared with reported experimental and theoretical results, and the mechanism for differences were discussed. This study provides insights into the thermal transport mechanisms of TaN, offering guidance for its application in advanced electronic and thermal management devices.

1. **Introduction**

Tantalum nitride (TaN) exhibits multifunctional characteristics, including high thermal conductivity, exceptional thermochemical stability, and tunable electronic properties[1-4]. These attributes make TaN a promising candidate for applications such as thermal management systems, high-temperature protective coatings, and advanced energy storage devices. In these applications, the level of thermal conductivity plays a critical role in determining performance and suitability[5-7].

The hexagonal θ-TaN phase (space group $P6_3/mmc$) and amorphous TaN (a-TaN) exhibit distinct differences in thermal conductivity and mechanical properties due to their differing structural configurations, leading to varied application scenarios. θ-TaN, with its relatively high thermal conductivity, holds potential for future use in transistor technologies[3,4]. In contrast, a-TaN, due to its structural and diffusion-resistant properties, is well-suited for use as a barrier layer in integrated circuit fabrication processes[8,9].

The hexagonal θ-TaN phase is predicted to exhibit promising thermal conductivity owing to its energy gap between Acoustic and optical branches. Kundu et al.[3,4] reported, based on solutions of the Boltzmann Transport Equation (BTE) and four-phonon scattering calculations, that the thermal conductivity of crystalline TaN could reach as high as 995 W/m-K under ambient conditions. Another phase, tetragonal tantalum nitride (t-TaN), was reported by Ding et al.[10] to have a thermal conductivity as high as 677 W/m-K at 300 K. To experimentally verify this high thermal conductivity, Liu et al.[11] synthesized θ-TaN through a phase transformation from ε-TaN powder. However, their measured thermal conductivity was only 47.5 W/m·K. Similarly, Lee et al.[12] prepared polycrystalline θ-TaN via high-pressure synthesis from the ε-phase and obtained a thermal conductivity of 90 W/m·K using time-domain thermoreflectance (TDTR) measurements. Both experimental results fall significantly short of the theoretical BTE prediction.

In contrast to the potentially high thermal conductivity of θ-TaN, the a-TaN phase typically exhibits significantly reduced thermal transport properties due to its disordered atomic configuration. Zhang et al.[13] prepared 100 nm-thick TaN thin films under varying $N_2$ flow rates and measured their thermal conductivity using TDTR, obtaining values in the range of 4-8 W/m·K. Similarly, Bozorg-Grayeli et al.[14] reported intrinsic thermal conductivities ranging from 3.0 to 3.4 W/m·K for TaN samples with thicknesses between 50 and 100 nm. However, previous experimental results often revealed that the synthesized amorphous TaN materials were actually mixtures of crystalline and amorphous phases, with relatively large feature sizes, thus failing to represent the application scenarios of amorphous TaN at the nanometer scale.

Previous studies on the thermal conductivity of both θ-TaN and a-TaN phases have lacked molecular dynamics (MD) simulations, primarily due to the absence of reliable empirical potentials. This methodological limitation can be effectively resolved by implementing machine learning potential (MLP) trained on ab initio datasets[15]. Ouyang et al.[16] trained a MLP for boron arsenide (BAs) and used it to simulate its thermal conductivity. Qin et al.[17] developed an MLP for a plastic crystal, a relatively complex crystalline material, and investigated the effect of strain on its thermal conductivity. Both studies demonstrate the effectiveness of MLP in accurately capturing thermal transport properties.

In this work, the values of thermal conductivity predictions for both θ-TaN and a-TaN phases are predicted by using MD with a trained MLP. Firstly, the DP model was trained for both crystalline and amorphous TaN, and the accuracy of the model was verified. Secondly, the thermal conductivity of bulk θ-TaN and a-TaN was calculated using the equilibrium molecular dynamics (EMD) method. Thirdly, the thermal conductivity of θ-TaN and a-TaN nanofilms was calculated using the non-equilibrium molecular dynamics (NEMD) method. Finally, the calculated results of thermal conductivity were compared and analyzed with the results obtained by different methods from previous studies.

## 2. Potential training and simulation methods

MLP have revolutionized computational materials science by enabling accurate and efficient thermal conductivity predictions across diverse material systems[18,19]. The DP model was trained using the active learning method (section 2.1) and its accuracy was verified (section 2.2). Subsequently, the details of how to simulate the thermal conductivity were explained (section 2.3).

### 2.1 Training of Deep learning potential

The DP model was trained using an iterative active learning workflow implemented in the DP-GEN software package[20]. This process involves generating structures, calculating their energies and forces via DFT, and training an ensemble of four DP models with identical parameters but different random seeds. The trained models are used to perform MD simulations, and deviations in forces among models are used to evaluate prediction reliability. Structures with high uncertainty are selected for additional DFT calculations, and the cycle repeats until the model reaches the target accuracy of 99%. More details can be found in Supply Materials (SM) I.

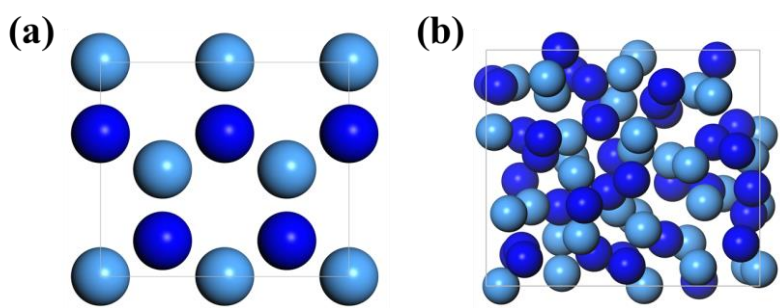

Fig.1 Strutures of (a) θ-TaN and (b) a-TaN.

The training process is grounded in high-fidelity DFT calculations and consistently validated through MD simulations to ensure the accuracy of the DP model. DFT data were generated using Vienna Ab initio Simulation Package (VASP)[21,22], with a plane-wave energy cutoff of 600 eV to ensure reliable convergence. The initial dataset was constructed from scaled and randomly displaced θ-TaN and a-TaN structures (structures shown in Fig. 1). All MD explorations were carried out in Large-scale

Atomic/Molecular Massively Parallel Simulator (LAMMPS)[23] using an ensemble of four independently trained DP models.

## 2.2 Accuracy of DP model

The DP model demonstrates agreement with DFT calculations in predicting both energies and atomic forces for configurations within the final training dataset, as shown in Fig. 3. Fig.3(a) and (b) shows the comparisons of energy per atom and atomic force between DP and DFT calculations for θ-TaN. Both the energy per atom and atomic force shows good agreement between DP and DFT. The results for a-TaN show that the error in the DP fitting is larger compared to θ-TaN. This is primarily due to the more complex atomic arrangements in amorphous materials. Previous studies also suggest that machine learning potential fitting for amorphous materials typically yields a slightly larger error compared with crystal materials[24-26]. Moreover, regarding its impact on thermal conductivity, phonon scattering is more pronounced in amorphous materials[27]. For such materials with low thermal conductivity, the fitting deviation generally leads to an underestimation of the thermal conductivity[28].

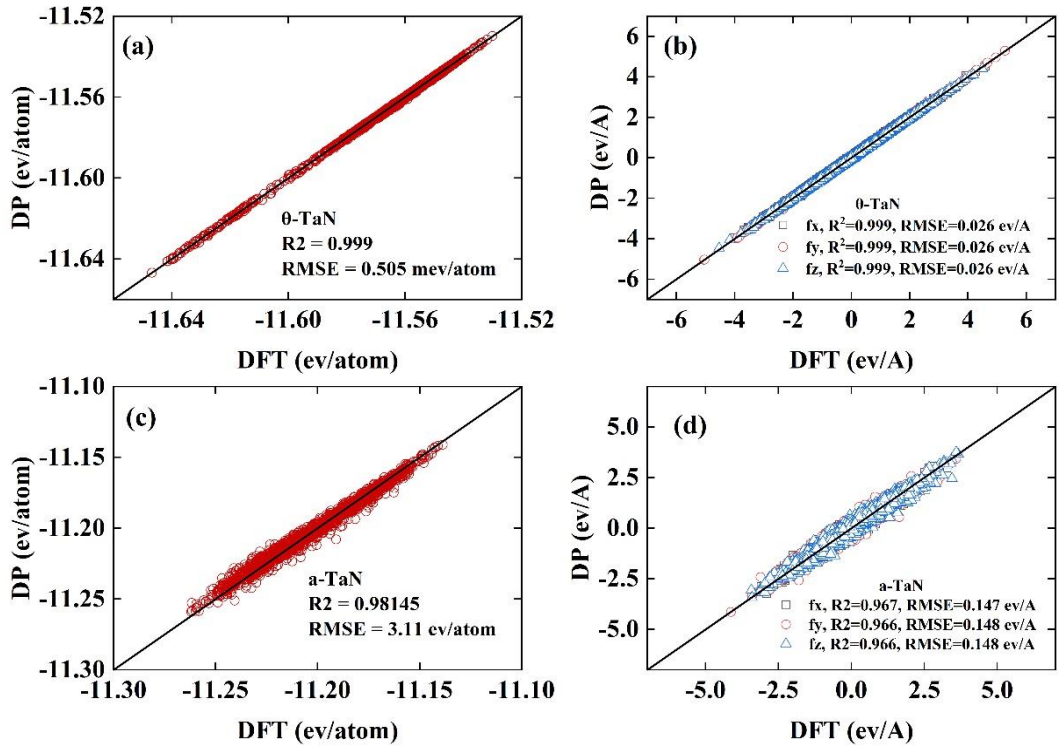

Fig.2 Comparisons of energy per atom (a) and atomic force (b) between DP and DFT

calculations for θ-TaN. Comparisons of energy per atom (c) and atomic force (d) between DP and DFT calculations for a-TaN. The RMSE is obtained by comparing the data between DP and DFT calculations.

The accuracy of the Deep Potential (DP) model is further validated through direct comparisons with DFT calculations. The phonon dispersion of θ-TaN was calculated using both the DP method and DFT, as shown in Fig 3(a). The acoustic branch, which governs heat conduction, showed excellent agreement between the DP method and DFT, thereby validating the accuracy of the DP(θ-TaN). Since phonon dispersion cannot be used for a-TaN, the energy results from scaling the volumes of both θ-TaN and a-TaN were calculated and compared between the DP method and DFT, as shown in Fig 3(b). The results showed good agreement, further verifying the accuracy of the DP(θ-TaN) and DP(a-TaN) models.

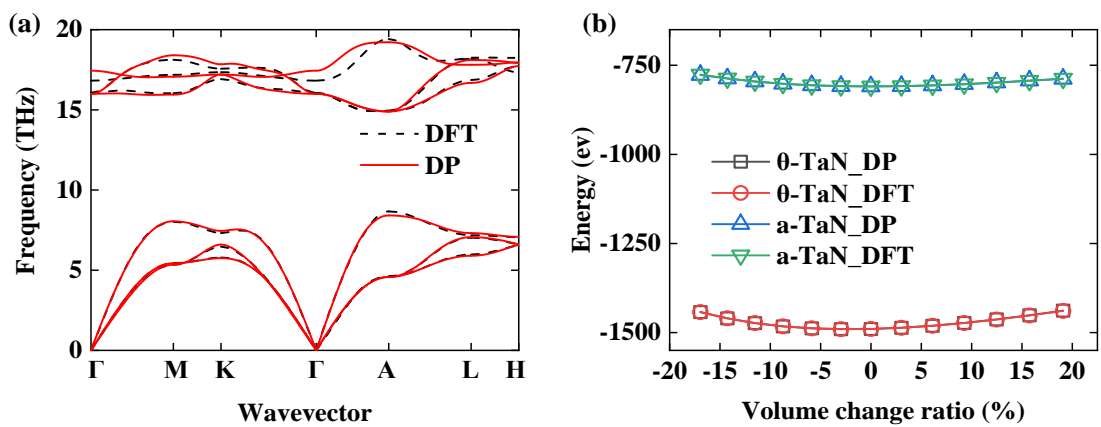

Fig. 3(a) The phonon dispersion relation of θ-TaN was compared with results obtained from DFT calculations. The acoustic branch, which primarily governs heat conduction, showed good agreement with the DFT results. (b) Energy changes from volume scaling of θ-TaN and a-TaN were compared with DFT results. The accuracy of DP(θ-TaN) and DP(a-TaN) models were verified.

## 2.3 Parameters setting of MD simulation of thermal conductivity

Both EMD and NEMD simulations were implemented using the LAMMPS with the DP models trained above. The equations of motion were integrated with the velocity

Verlet method using a time step of 0.5 fs. Prior to the calculation of thermal conductivity of both EMD and NEMD, the structures underwent a NPT ensemble (constant number of atoms, pressure, and temperature) at 300 K and 0 bar for 0.5 ns.

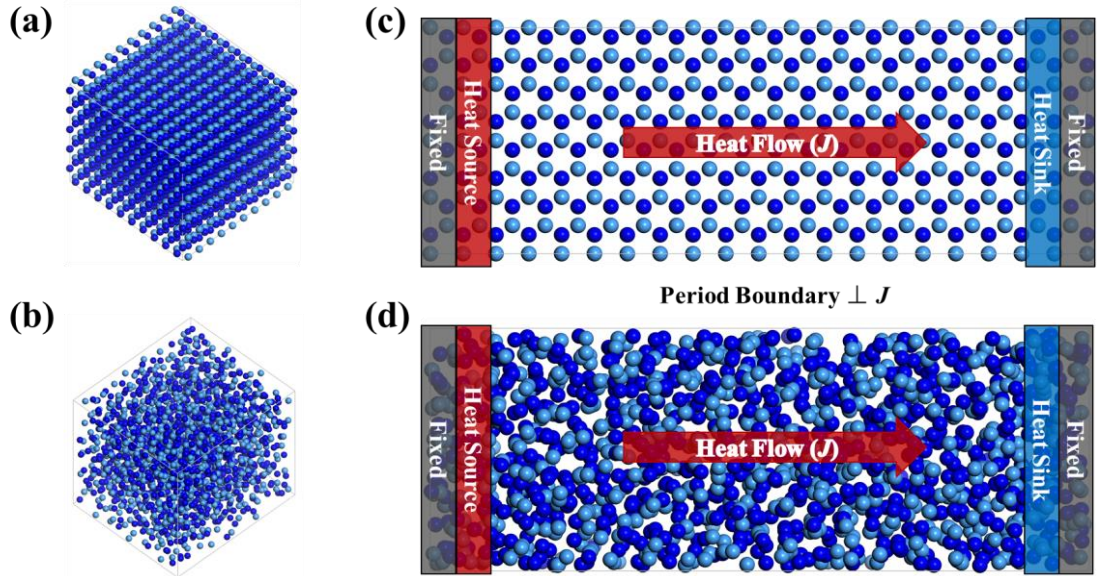

Fig. 4 Structures used in MD simulations: (a) and (b) show the configurations of θ-TaN and a-TaN used in EMD simulations, where periodic boundary conditions were applied in all three directions. (c) and (d) present the corresponding structures used in NEMD simulations. In NEMD, fixed boundary conditions and Langevin thermostats were applied along the heat flow direction, while periodic boundary conditions were imposed in the directions perpendicular to heat flow to simulate an infinitely large film.

The thermal transport properties of bulk θ-TaN and a-TaN were investigated using the EMD simulation method with three-dimensional periodic boundary conditions. The simulation protocol consisted of three sequential stages: (1) Structural relaxation was initially performed in the isothermal-isobaric (NPT) ensemble at 300 K and 0 bar for 500 ps to achieve thermodynamic equilibrium. (2) Subsequently, the system was equilibrated in the canonical (NVT) ensemble for 200 ps to stabilize the temperature at 300 K. (3) Finally, production runs were conducted in the microcanonical (NVE) ensemble over 4 ns for data acquisition. Special attention was given to the calculation of heat current autocorrelation functions (HCACF), where a sampling interval of 5 simulation steps was implemented. The correlation time parameters were empirically determined as 1000 ps for θ-TaN and 20 ps for a-TaN, reflecting their distinct structural characteristics.

The thermal transport characteristics of a-TaN thin films with varying lengths were investigated using NEMD simulations. The computational domain employed fixed boundary conditions along the heat flux direction (x-axis) and periodic boundaries in transverse directions (y- and z-axes). Temperature gradients were imposed using Langevin thermostats: a 310 K thermal reservoir was applied to atomic layers 3–6, while layers (N–6) to (N–3) were coupled to a 290 K bath (N=40 total layers). The thermostat damping coefficient was optimized to 50 times the simulation timestep (25 fs), balancing efficient phonon scattering with minimal artificial thermal resistance. Boundary layers (1–2 and N–1 to N) were positionally constrained to eliminate rigid-body motion. The simulation protocol comprised two phases: (1) A 2 ns NVE equilibration stage to establish steady-state thermal transport under Langevin bath coupling, followed by (2) a 5 ns (for θ-TaN) or 2.5 ns (for a-TaN) production phase in the NVE ensemble for time-averaged measurements of temperature profiles and heat flux. Statistical reliability was ensured through five independent simulations with randomized initial atomic velocities. Full methodological details, including EMD and NEMD, are provided in SM II and SM III.

## 3. Results and discussions
### 3.1 Bulk θ-TaN and a-TaN by EMD

The HCACF and its integral along all directions of θ-TaN and a-TaN at 300 K are shown in Fig. 5. The HCACF initially exhibits a high value, indicating strong temporal correlation of the heat current at short timescales. It then decays rapidly, reflecting a quick loss of correlation within the system. At longer timescales, slight oscillations appear, typically attributed to phonon scattering or local structural vibrations. By integrating the HCACF over time and applying the Green–Kubo formula, the thermal conductivity of the material can be determined.

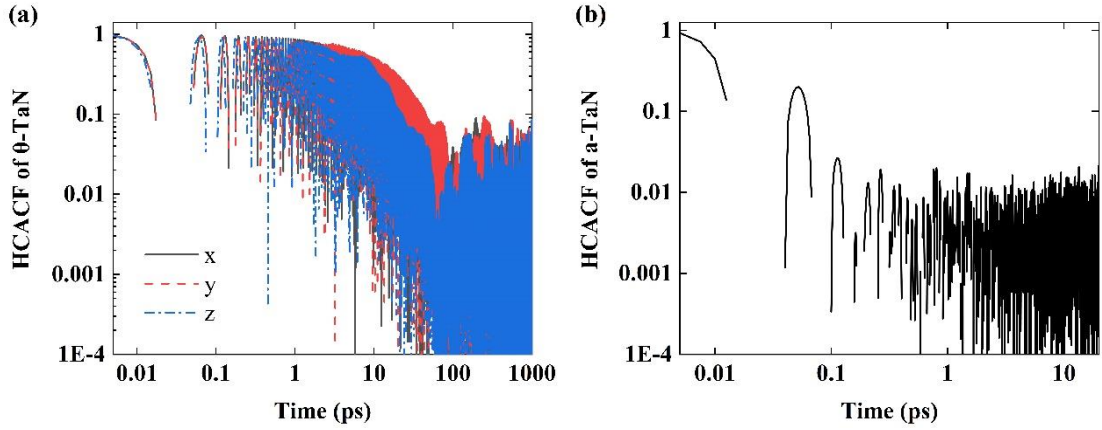

Fig.5 HCACF of (a) θ-TaN and (b) a-TaN. The HCACF exhibits a high initial value, indicating strong short-time heat current correlation, followed by rapid decay due to fast loss of correlation. At longer timescales, slight oscillations appear, typically regarded as statistical noise. When integrating the HCACF to calculate thermal conductivity, care should be taken to minimize the influence of this noise.

The anisotropic thermal conductivity of bulk θ-TaN is demonstrated in Fig. 5(a)-(c), showing distinct values along the crystallographic [100] (x), [010] (y), and [001] (z) directions, respectively. The grey lines represent 25 independent simulations conducted at 300 K. The red solid line shows the averaged thermal conductivity across these simulations, while the red shaded region indicates the corresponding uncertainty range. A convergent trend in the averaged values is observed for θ-TaN as the integration time becomes sufficiently long. The final thermal conductivities along the x-, y-, and z-directions were obtained by averaging the values over the last 50 ps. The standard error (SE) was used to estimate the uncertainty, yielding values of 530±90, 430±90, and 470±90 W/m·K, respectively.

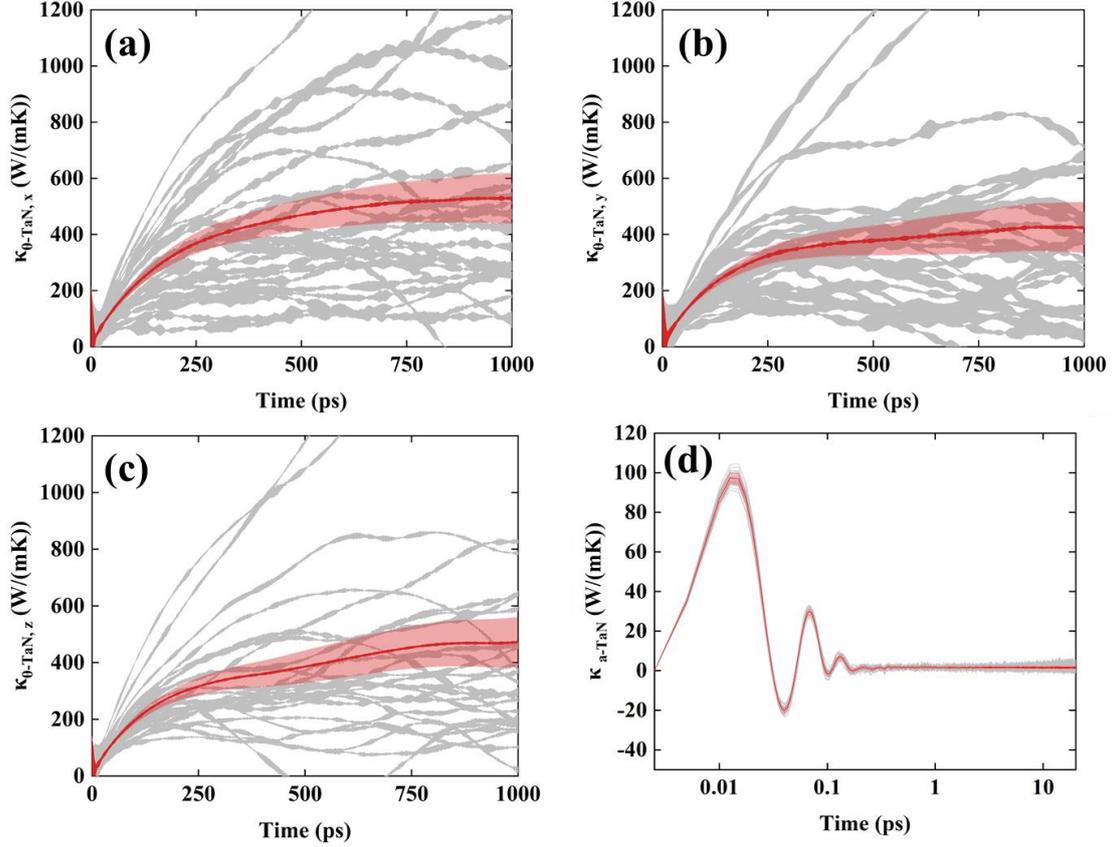

Fig.6 Thermal conductivity of TaN calculated using EMD simulations: (a) x-axis of θ-TaN, (b) y-axis of θ-TaN, (c) z-axis of θ-TaN, and (d) a-TaN. The grey lines represent 25/10 independent simulations for θ-TaN/a-TaN conducted at 300 K. The red solid line shows the averaged thermal conductivity across these simulations, while the red shaded region indicates the corresponding uncertainty range. For θ-TaN, the final thermal conductivity values are obtained by averaging the last 50 ps of the simulation; for a-TaN, the average is taken over the last 1 ps. The final thermal conductivity values for the x, y, and z directions of θ-TaN and for a-TaN are 530 ± 90, 430 ± 90, 470 ± 90 and 1.6 ± 0.2 W/m·K, respectively.

The isotropic thermal conductivity of bulk θ-TaN is demonstrated in Fig. 6(d). The grey lines represent 10 independent simulations conducted at 300 K. The red solid line shows the averaged thermal conductivity across these simulations, while the red shaded region indicates the corresponding uncertainty range. A convergent trend in the averaged values is observed for a-TaN as the integration time becomes sufficiently long. The final thermal conductivity was determined by averaging the values over the last 1 ps. The SE was used to quantify the uncertainty, yielding a value of 1.6 ± 0.2 W/m·K.

The thermal conductivity of bulk amorphous θ-TaN and a-TaN is compared with previously reported values, as summarized in Table 1. Previous studies have lacked MD simulations. This work fills that gap and provides comparisons with prior theoretical calculations and experimental results for θ-TaN, as well as with experimental measurements for a-TaN. Due to structural differences between the simulated and experimentally measured samples, along with methodological differences between simulations and theoretical calculations, some differences in results are observed in the results.

Table 1 Thermal conductivity of θ-TaN ($\kappa_{\theta-TaN}$) and a-TaN ($\kappa_{a-TaN}$) compared with previous works

| Structures & Reference | Method | Temperature (K) | Thermal Conductivity (W/m·K) |
|---|---|---|---|
| **θ-TaN (This work)** | **EMD (Simulation)** | **300** | **X (a-axis): 530±90<br>Y: 430±90<br>Z (c-axis): 470±90** |
| θ-TaN[3,4] | BTE+3phonon (Calculation) | 300 | a-axis (X): ≈2000 |
| θ-TaN[3,4] | BTE+4phonon (Calculation) | 300 | a-axis (X): 995<br>c-axis (Z): 820 |
| θ-TaN[11] | TDTR (Experiment) | 300 | 47.5 |
| θ-TaN[12] | TDTR (Experiment) | 300 | 90 |
| **a-TaN (This work)** | **EMD (Simulation)** | 300 | **1.6 ± 0.2** |
| a-TaN & θ-TaN[13] | TDTR | 300 | 4.0 – 8.0 |
| (100nm) | (Experiment) | | |
| a-TaN & θ-TaN[14] (50-100nm) | TDTR (Experiment) | 300 | 3.0 - 3.4 |

For θ-TaN, the results of the EMD simulation fall between the theoretical calculation results of BTE and the experimental measurement results. Compared with the results obtained from BTE calculations, the thermal conductivity predicted by EMD simulations is generally lower. This discrepancy arises from several factors. First, BTE calculations currently consider up to four-phonon scattering processes. While the inclusion of four-phonon interactions significantly reduces the predicted thermal conductivity compared to three-phonon calculations, higher-order phonon interactions—potentially relevant in strongly anharmonic materials—are still neglected [29,30]. This limitation may lead to an overestimation of thermal conductivity in BTE results. Additionally, the experimentally measured thermal conductivity is often lower than the simulation results due to structural differences. Experimentally synthesized crystals are typically polycrystalline, and phonon scattering at grain boundaries plays a significant role. The presence of such boundaries reduces phonon mean free paths and thus lowers the observed thermal conductivity, compared to the idealized structures modeled in simulations. Furthermore, in the measurement of thermal conductivity, TDTR still employs the Fourier heat conduction model. High thermal conductivity materials typically exhibit long phonon mean free path (MFP). When the thermal excitation is confined to a small region, such as the localized heating induced by a focused laser spot in TDTR experiments, nonlocal and partially ballistic heat transport effects can arise, making Fourier's law inadequate. As a result, the thermal conductivity obtained from Fourier-based models is possibly systematically underestimated[31,32].

For a-TaN, the thermal conductivity obtained from EMD simulations is relatively lower than the experimental values. This discrepancy is primarily attributed to structural differences. The experimental samples often consist of a mixture of amorphous and crystalline phases, whereas the simulations are based on fully amorphous structures. Since crystalline materials generally exhibit higher thermal conductivity than amorphous ones, the presence of crystalline regions in the experimental samples likely leads to an overestimation of the thermal conductivity of purely a-TaN. This explains why the EMD results are lower than the experimental measurements.

### 3.2 θ-TaN and a-TaN nanofilms by NEMD

The temperature profiles and energy exchange data from the Langevin thermostat for θ-TaN and a-TaN thin films, obtained via NEMD simulations, are presented in Fig. 7. Both materials exhibit stable temperature gradients, indicating steady-state heat transport. The energy accumulated by the heat sink and released by the heat source corresponds to the net heat flux across the system. Based on this, the thermal conductivity can be determined using Fourier's law of heat conduction:

$$\kappa_f = -\frac{J}{S \cdot \nabla T} \quad (1)$$

$$\kappa_d = -\frac{JL}{S \cdot \Delta T} \quad (2)$$

where $J$ represents the heat flux, $S$ denotes the cross-sectional area of heat transfer, and $\nabla T$ stands for the temperature gradient, $\Delta T$ stands for the temperature difference the endpoints of the heat transfer zone divided by its length $L$.

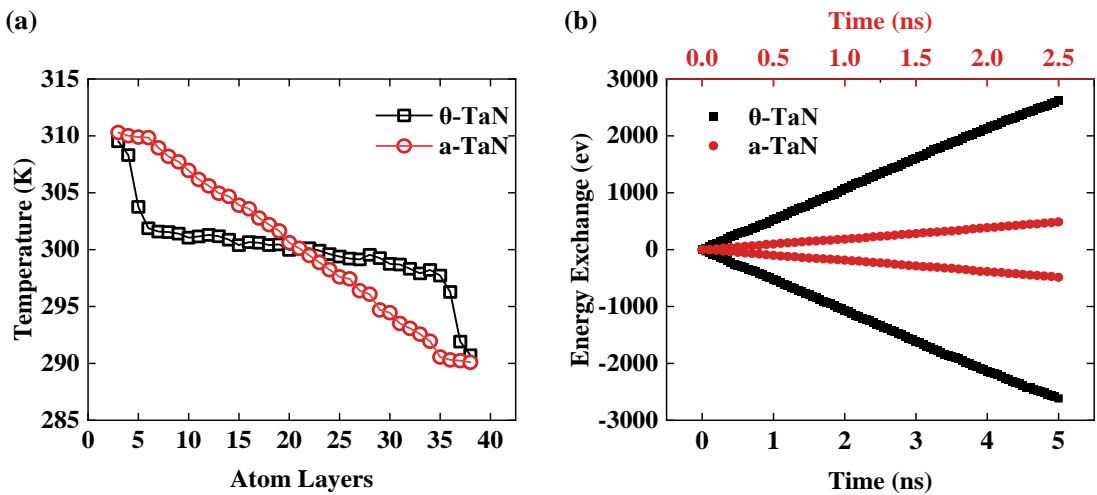

Fig.7 The (a) temperature profiles and (b) energy exchange of the Langevin thermostat for θ-TaN and a-TaN thin films. Two methods were employed to calculate the temperature gradient: (I) linear fitting of the temperature profile in the intermediate heat transfer region ($\nabla T_f$), and (II) direct calculation using the temperature difference ($\Delta T$) between the endpoints of the heat transfer zone divided by its length ($L$). Based on these gradients, the corresponding thermal conductivities $\kappa_f$ and $\kappa_d$ were obtained.

When applying Fourier's law to calculate thermal conductivity, careful treatment of the temperature gradient is essential. For a-TaN, the temperature gradient can be reliably obtained via linear fitting in the central heat transfer region, as shown in Fig. 7a. In contrast, θ-TaN, which possesses a relatively long phonon mean free path, presents additional challenges. At the nanoscale, non-negligible ballistic transport near the boundaries causes deviations from a purely diffusive regime, making a straightforward linear fit insufficient to accurately characterize the gradient[33,34]. To address this, two methods were employed for θ-TaN: I. linear fitting $\nabla T_f$ of the temperature profile in the intermediate heat transfer region, and II. calculating the temperature difference $\Delta T$ across the heat transfer zone endpoints and dividing it by the corresponding length $L$. Then thermal conductivity of both $\kappa_f$ and $\kappa_d$ can be calculated.

Significant size-dependent behavior is observed in both $\kappa_f$ and $\kappa_d$ for θ-TaN thin films, as calculated via NEMD simulations and shown in Fig. 8(a). Within the length range of 5.0–15.0 nm, both $\kappa_f$ and $\kappa_d$ increase rapidly. The relationship between thermal conductivity and film length is fitted using the formula $\kappa \propto L^\beta$.

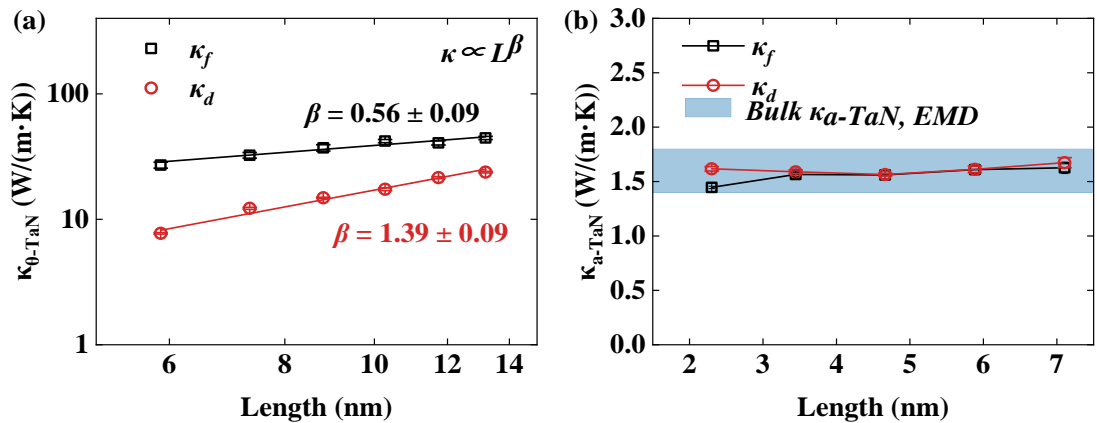

Fig.8 (a) Thermal conductivity of θ-TaN nanofilms as a function of film size, calculated using NEMD simulations at 300K. Significant size-dependent behavior is observed in both values $\kappa_{fitting}$ and $\kappa_{direct}$. The relationship between thermal conductivity and film length is fitted using the formula $\kappa \propto L^\beta$. (b) Thermal conductivity of a-TaN films as a function of film thickness, calculated using NEMD simulations at 300 K. Across the thickness range of 1.5–7.5 nm, the thermal conductivity remains nearly constant at approximately 1.5–1.6 W/m·K, showing no evident size dependence. These results are consistent with those obtained from EMD calculations (indicated by the blue shaded region).

The thermal conductivity of a-TaN thin films, calculated via NEMD simulations, is presented in Fig. 8(b). Across the thickness range of 1.5–7.5 nm, the thermal conductivity remains nearly constant at approximately 1.5–1.6 W/m·K, consistent with the values obtained from EMD calculations. Furthermore, the absence of significant variation with thickness suggests that size effects on the thermal conductivity of a-TaN are negligible within this range.

## 4. Conclusion

In summary, the bulk and nanofilm thermal conductivities of both θ-TaN and a-TaN were investigated using MD simulations with DP models. Two DP models were trained for θ-TaN and a-TaN using an active learning approach. The accuracy of the trained models was validated by comparisons of per-atom energy, atomic forces, energy–volume relationships, and phonon dispersion (for θ-TaN only). The DP models demonstrated agreement with DFT results, confirming their reliability for predicting thermal conductivity in both crystalline and amorphous phases.

The calculated bulk thermal conductivities were compared with available experimental measurements and BTE results. For θ-TaN, the maximum directional thermal conductivity predicted by EMD simulations reaches 530 ± 90 W/m·K, which is significantly higher than both the experimental measurements (47.5 and 90 W/m·K) and the theoretical values obtained via the BTE with four-phonon scattering included

(995 W/m·K). This discrepancy can be primarily attributed to the presence of grain boundaries and defects in experimentally synthesized materials, which reduce phonon transport efficiency. Furthermore, both experimental and theoretical approaches have inherent limitations: experimental measurements often rely on the Fourier heat conduction model, which may underestimate thermal conductivity at nanoscale due to ballistic transportation, while BTE-based calculations typically neglect higher-order phonon scattering and anharmonic effects beyond four-phonon interactions. For a-TaN, the simulated bulk thermal conductivity ($1.6 \pm 0.2$ W/m·K) is lower than the experimentally measured values (3.0–3.4 W/m·K and 4.0–8.0 W/m·K), primarily due to the presence of crystalline phases within the experimental samples. These crystalline components possess higher intrinsic thermal conductivity, thereby increasing the overall thermal conductivity of the amorphous–crystalline mixtures observed in experiments.

Additionally, in the study of nanofilm thermal transport, θ-TaN exhibited a pronounced size effect, whereas a-TaN showed negligible size dependence. In θ-TaN, a pronounced size dependence was observed, which is attributed to its crystalline structure and the presence of long mean free path phonons that undergo significant boundary scattering as film thickness decreases. In contrast, a-TaN exhibited negligible size dependence due to its inherently localized vibrational modes and the absence of long-range phonon transport, resulting in thermal conductivity that remains largely insensitive to variations in film thickness.

This study not only reveals the distinct thermal transport behaviors of θ-phase and amorphous TaN, but also provides important guidance for their respective applications in nanoscale film technologies. The relatively high thermal conductivity of crystalline θ-TaN suggests its potential as a thermal management material or even as an alternative to conventional semiconductors in specific applications. In contrast, amorphous TaN, often employed as an ultra-thin diffusion barrier in device fabrication, exhibits weak size dependence in its thermal conductivity, which ensures stable thermal performance even at nanometer-scale thicknesses.

## ACKNOWLEDGMENTS

The work was carried out at the National Supercomputer Center in Tianjin, and the calculations were performed on TianHe-HPC.

## DATA AVAILABILITY

The data that support the findings of this study are available from the corresponding author upon reasonable request.